# SRMCA: A *S*calable and Resilient *R*eal-time *M*ulti-party *C*ommunication *A*rchitecture

Asit Chakraborti, Syed Obaid Amin, Aytac Azgin, *Member, IEEE,*
Ravishankar Ravindran, *Senior Member, IEEE,* and Guo-Qiang Wang

***Abstract*—This paper proposes a scalable and resilient real-time multi-party communication architecture for the delivery of mixed media streams, for which *content centric networking*, with its intelligent network layer, is chosen for implementation to address the shortcomings of the current `IP`-based Internet architecture. Content centric networking (`CCN`) represents one of the major proposals targeting future Internet architecture, and is typically optimized for non-realtime content delivery. Proposed research in this paper addresses the architectural challenges for large-scale deployment of `CCN` to serve real-time applications from the perspective of a multi-party video conferencing framework, with the necessary architectural components to support application- and system-level objectives, in regards to quality of experience, resource utilization, and scalability. We present an in-depth analysis of the proposed architecture by first providing an analytical driven study for the system, and then demonstrating its performance by emulating the architecture over a test-bed with a large number of participants involved in many-to-many communication sessions. The results suggest that the proposed architecture can scale well above 50 participants without incurring significant penalty in signaling, communication, and computing overheads, with capability to support 100 or more participants.**

## I. Introduction

According to a recent service adoption forecast, video conferencing is expected to represent a significant portion of the video traffic, while growing at an annual rate of ≈ 52% [1]. Earlier studies have shown scalability limitations for `IP`-based conferencing systems (*e.g.*, [2], [3]), which are typically caused by a lack of a general multicast support in the current Internet.

Current `IP`-based conferencing systems can typically be classified as *peer-to-peer* (`P2P`), *client-server*, or *hybrid* solutions, depending on how data flows among participants. While a `P2P` approach uses direct transport sessions, a client-server approach uses server(s) as relay node(s). In the case of a hybrid architecture, end hosts connect to one of the available servers, and the servers inter-connect among themselves in a `P2P` manner creating a relay network at the application level. Regardless of the type of architecture used for conferencing, scalability problems triggered by the `IP`'s deficiencies continue to persist.

To exemplify the impact of such problems, we tested the performance of client-server based solutions using `ns3` for the `Rocketfuel`-1755 topology (consisting of 87 nodes and 322 links), and show the results (average of 15 runs with different random seeds) on bandwidth use in Figure 1 for various client-server scenarios (by distributing clients and servers randomly in the topology, and assigning clients to the closest server). In Figure 1(a) we show the cumulative distribution function plot for the bandwidth utilization in the network assuming a single-server case, and demonstrate how the distribution for bandwidth utilization (normalized with respect to single stream rate, *i.e.*, we divided the actual link rates with the single stream rate) is affected as the number of clients increases. We observe that as the number of users increases, more and more links are being utilized at higher rates.[1] For instance, with just 32 clients, 5% of the network links observe bandwidth utilization of more than 100 streams. Such use can become prohibitive as more and more streams are activated. Similarly, in Figure 1(b) we show the impact of introducing additional servers to the system, to help with distributing the network load by creating a server multicast before delivering content to clients from the servers. We observe noticeable improvement in performance as we introduce additional servers, but the improvements, for the given topology, reach a convergence point around 5 servers, thereby limiting the performance improvements to ≈ 40%.[2] These results clearly illustrate the limits of an `IP`-

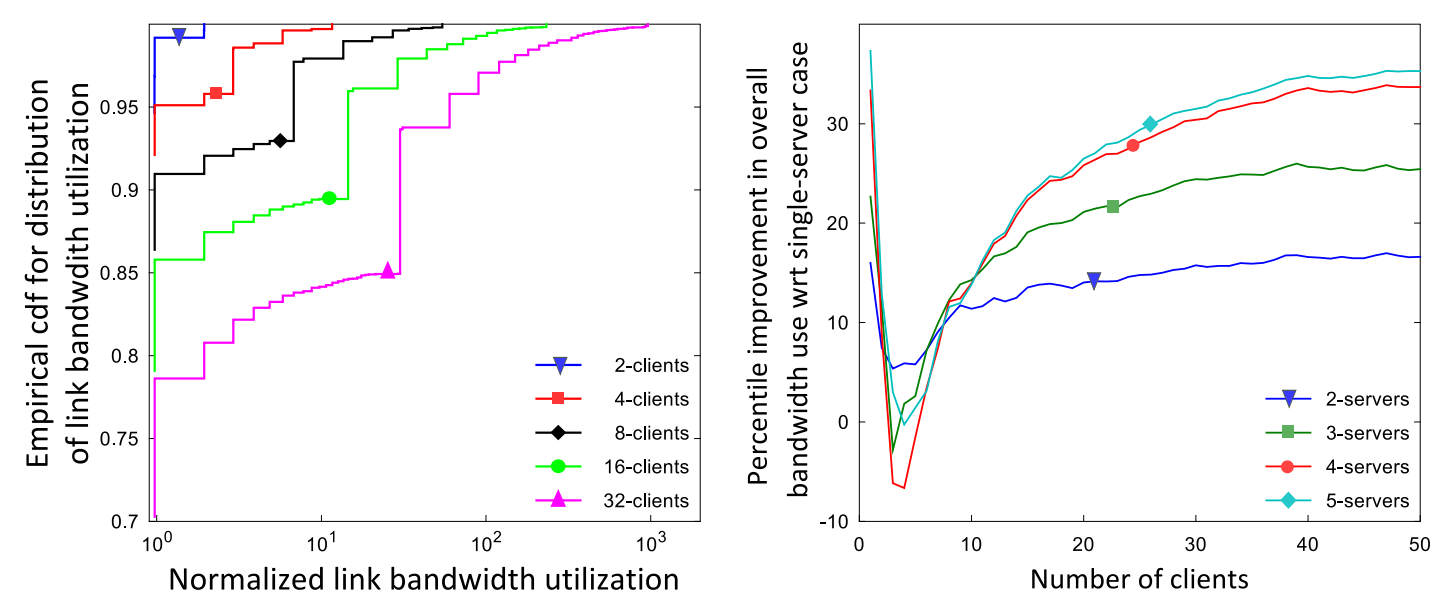

(a) Link utilization for single-server case.

(b) Impact of introducing additional servers.

Fig. 1. Network link bandwidth utilization for client-server based architectures.

The authors are with Huawei Research Center, Santa Clara, CA, 95050 USA e-mail(s): {asit.chakraborti, obaid.amin, aytac.azgin, ravi.ravindran, gq.wang}@huawei.com.

[1] Specifically, for the single server case, overall network bandwidth utilization increases at a rate $\propto n^{1.947}$ and average bandwidth utilization over active links increases at a rate $\propto n^{1.346}$, where $n$ represents the number of participants.

[2] For the 5-server case, overall network bandwidth utilization increases at a rate $\propto n^{1.872}$ and average bandwidth utilization over active links increases at a rate $\propto n^{1.183}$, with $n$ representing number of participants.

based architecture with limited perceived improvements even as we increase the number of servers; as IP does not support multicast natively, we need to emulate multicast atop unicast, in which case the multicast architecture continues to rely on unicast-based forwarding from the server to the clients.

To address these problems and others, future networking architectures have been considered to replace the current IP, which would be highly beneficial to any content delivery system, including conferencing systems [4]. For instance, we can use information-centric networking (ICN) to enable several desirable features to support conferencing applications, which include features like in-network multicast supporting many-to-many communications and in-network caching and computing [5]. These benefits are visible even in an overlay deployment mode, as demonstrated through this work. Among the existing proposals for ICN, a popular architecture that has gained significant attention (in both academia and industry) is the content centric networking architecture (CCN, or named data networking, NDN[3]) [6]. It aims to replace the current IP's host-centric design with a content-centric one with the named content becoming the principle entity in networking.

CCN utilizes a *pull-based* content delivery framework with Interest/Data primitives, which allows the architecture to efficiently support non-realtime services like video streaming [7], [8]. However, the desire to meet the stringent latency requirements of real-time applications, which seek *push-based* network services, bring forth additional challenges that need to be carefully addressed. For instance, in [9], the authors propose request (or Interest) pipelining to handle real-time content fetching, and in doing so, demonstrate the applicability of CCN to support voice services. In [10], the authors address similar challenges in NDN to support video conferencing, and successfully demonstrate their results over the NDN testbed [11]. However, for these architectures, no large scale performance results in regards to quality of experience (QoE) for multi-party and interactive video or voice content have been reported.

We can summarize the main challenges in adopting a CCN-based architecture to implement a multi-party conferencing system as follows:

- The use of many-to-many communications within conferencing requires service-level optimizations in the utilization of computing resources of the network to allow for scaling to support a large number of participants;
- Random join or leave by a participant requires actively syncing a producer's state with each consumer, while taking into account the QoE requirements for the end hosts (or user entities, UEs);
- As consumers can become out-of-sync with producers during an active session, we need mechanisms to handle such scenarios to re-sync user states.

In this paper, we address these challenges in the context of CCN using a real-time notification framework that runs over the service-friendly CCN transport and relies on in-network services to reduce the overall complexity at the end hosts. Proposed scalable video conferencing architecture is capable of (*i*) handling multiple simultaneous conference instances, and (*ii*) allowing participants to quickly synchronize with the participant state after random joins or transient disconnects. We address the scalability challenges that are discovered in our earlier research [12] through decoupling notification framework used for name-syncing from content fetching. We realize our solution over a programmable CCN transport that is controlled by an ICN aware Network Function Virtualization/Software Defined Networking (NFV/SDN) equivalent framework (for details, see [13]), which allows it to be amenable to edge deployment, and thereby addressing the ICN's near-term deployment challenge. The resulting architecture is shown to be capable of supporting up to or more than 100 participants, depending on the evaluation context and the system parameters.

In short, we can summarize our contributions as follows:

- We address the challenges of realizing a large-scale real-time multi-party communication framework by proposing an ICN based architecture that takes advantage of ICN capabilities at the network layer (such as multicasting) and combines it with an edge service framework that not only allows for efficient multicasting of multi-party content but also allows for offloading complex system components from user entities to the network to achieve a scalable and resilient solution.
- We present an in-depth analysis of the proposed architecture, both at a component level and in regards to user- and system-level performance to evaluate its unique features and understand its operation. We utilize a large-scale performance study of our solution to investigate its impact on two uniquely important real-time multi-party communication scenarios, and demonstrate its robustness in supporting both cases while addressing the potential bottlenecks that may prove crucial for future adoption of similar schemes.

The remainder of the paper is organized as follows. Section II discusses our conferencing architecture, along with the notification framework that is used to address the issue of name-syncing among dynamic conference participants. In Section III, an analytical framework is introduced to understand the operational basics of the proposed solution, with the objective of identifying the parameters and their settings affecting the overall quality of our solution. In Section IV, we evaluate the performance of the proposed architecture. We first describe the general settings for our testbed, including the traffic models and the application settings, then we study the performance of our solution in regards to expected savings in bandwidth use and compute resources, participant scalability, and recovery from temporary disruptions. We discuss the impact of these performance measures on end-to-end application level latency and throughput per the real-time multimedia QoE requirements.We summarize the existing research on real-time communications in information-centric networks in Section V and conclude our paper in Section VI.

[3]Hereafter, we use CCN and NDN interchangeably.

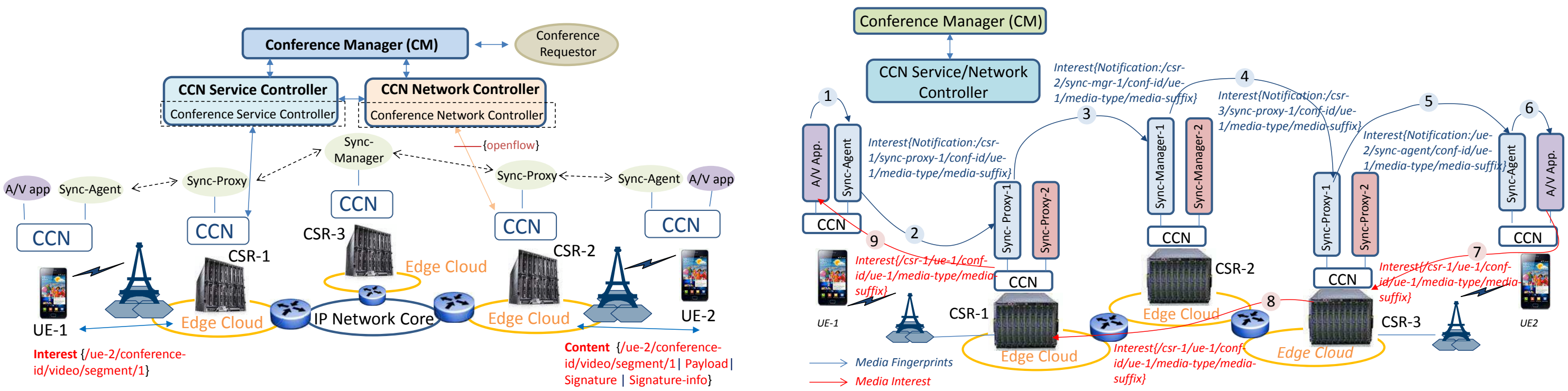

(a) Conferencing framework

(b) Notification framework

Fig. 2. Architectural diagrams for the video conferencing and notification frameworks

## II. Conference System Design

We show in Figure 2(a) the proposed conferencing architecture, which is illustrated in the context of an overlaid `CCN` deployment that is realized at the network edge. Here, a `CCN Service Router` (`CSR`) is an edge `CCN` node that is capable of supporting (or aiding) applications, in our case the conferencing service, and providing consumers with service discovery primitives to enable the discovery of a `CSR` and other services enabled in the network. The placements of the services in a `CCN` network is managed by the service orchestration, which can be integrated into an `SDN`/`NFV`-based framework for `IP`. This paper proposes a conferencing service architecture which focuses on the participant `QoE` and service scalability given a `CCN`/`NDN`-based infrastructure, rather than complimentary optimizations possible at the service orchestration level.

In the proposed architecture, at the application level, each `UE` connects to its nearest `CSR` either using local service discovery or through local configuration. `UE` runs two important applications: (*i*) audio/video (`A/V`) conferencing, with which the `UE` interacts, and (*ii*) `Sync-Agent`, which is part of the name-sync framework that helps `UE`s synchronize namespaces with other participants. `Sync-Agent` is responsible for receiving updates–in regards to the latest conference state (*i.e.*, namespaces for the `A/V` applications run by remote participants that belong to the same conference instance)–from the `Sync-Proxy` hosted on the `CSR`.

`Sync-Manager` is a central network-service instance that communicates with all the `Sync-Proxy` instances hosted on the `CSR`s to address the name-sync issue. For simplicity, we consider the name-sync framework to operate in a hub-n-spoke manner, for which the details are provided shortly, in Section II-B. Both the `Sync-Proxy` and the `Sync-Manager` services are implemented as virtual machines, allowing for on-demand provisioning and scaling depending on the active conferencing sessions.

`Conference Manager` (`CM`) is a logically centralized service controller that handles the following tasks:

- During conference provisioning (*i.e.*, allocating resources for the conference session), `CM` handles resource management, *i.e.*, depending on the expected system load (*e.g.*, bandwidth, computing), `CM` determines the number of `Sync-Proxy`s, their locations, and the location, where the `Sync-Manager` has to be provisioned;
- `CM` also manages the knowledge regarding the current conference sessions, including the set of active participants for each conference;
- `CM` can also handle the security functions, such as exchanging group keys to secure the conference session. In the data-centric security model considered for `CCN`, group keys are used to establish provenance, integrity and confidentiality of the producer's content.

`CCN Service/Network Controllers` implement `CCN`-driven `NFV` and `SDN` functions. Specifically, `CCN Service Controller` manages the compute virtualization of the `CSR` resources, by monitoring the compute resource usage in real-time and provisioning service functions accordingly at the `CSR`s (*i.e.*, `Sync-Proxy` and `Sync-Manager`). `Conference Service Controller` monitors these service functions and reacts accordingly to certain events (*i.e.*, failure or request to provision more instances from the `CM`). `CCN Network Controller` virtualizes `CCN` transport to several applications (such as `Conference Network Controller`) to dynamically provision the forwarding tables (*i.e.*, `FIB`) based on the application requirements. `Conference Network Controller` executes the conference logic in real-time to perform the following tasks: (*i*) managing virtual topologies for each conference session, (*ii*) handling `UE` join/leave events for conference sessions, (*iii*) provisioning the `FIB` for inter-connecting `CSR`s to `UE`s, and (*iv*) inter-connecting the service functions. Our proposed solution leverages `OpenStack` [14] and `ONOS` [15] to enable these functions.

Next, we discuss the `CCN` naming ontology adopted in our solution, which helps understanding the name-sync framework.

### *A. Naming*

In the proposed conference framework, we define two namespaces: (*i*) *content* (or *data*) *name*, which is used for user data exchange, and (*ii*) *notification name*, which is used for the name-sync protocol.

Data names correspond to the media content generated by the UE, and they are prefixed by the CSR, which the UE connects to, and follows the basic format of */<CSR-Gateway-ID>/<Conf-Session-ID>/<UE-ID>/<Media-Type>/<[Media-Suffix]>*. In the given naming format, CSR-Gateway-ID identifies the ID of the CSR node that the UE uses as the CCN gateway. Conf-Session-ID is a unique ID corresponding to the given conference session. UE-ID identifies the producer application. Media-Type identifies the type of content, which is either audio, video, or text. Media-Suffix includes Media-Type specific sub-components.[4] Note that, proposed architecture is not limited to static hosts. To support mobile UEs using a solution based on [16], naming convention would require topology independence, in which case, an alternative naming such as */<UE-ID>/<Conf-Session-ID>/<Media-Type>* can be used to help with content mobility. However, for the current research, to understand the capabilities offered by the proposed architecture, UEs are assumed to be static, which allowed us to use the former naming schema.

Notification names correspond to name-sync messages generated by the Sync-Agent and forwarded to the Sync-Proxy (which then forwards the messages to the Sync-Manager), and follow the convention of */<CSR-Gateway-ID>/<Service-Function-ID>/<[Fingerprint]>*. Each notification is a point-to-point message between Sync-Agent, Sync-Proxy, and Sync-Manager, and carries the fingerprint of the latest namespace used by the producer's A/V application. Fingerprints are derived from the application components of a user's data name that uniquely identifies the Conf-Session-ID, UE-ID, Media-Type, and Media-Suffix.

### *B. Notification (or Name-Sync) Framework*

Notification framework is used to sync the state of a producer's media namespace with participants in a given conference session's context, and is application independent (*i.e.*, leveraged by any application). Figure 2(b) shows the notification framework, which consists of three main functional components: Sync-Agent, Sync-Proxy, and Sync-Manager.

Sync-Agent is hosted at the UE and interacts with UE's producer application, which has multiple media streams to synchronize. The frequency of updates between producer and the Sync-Agent is media dependent.

Sync-Proxy is hosted at the CSR and provisioned by the CCN Service Controller to concurrently handle multiple conference sessions. Sync-Proxy participates in conference management by notifying the Conference Network Controller whenever a participant registers or deregisters with it. As shown in Figure 2(b), multiple Sync-Proxys can be provisioned to serve different conference sessions. Mapping between a UE's Sync-Agent and the CSR's Sync-Proxy is established during service discovery, upon which the Sync-Agent registers itself to a conference session, which is then used by the Sync-Proxy to update the Conference Network Controller. Sync-Proxy differentiates between the notifications targeting different conference sessions using the Conf-Session-ID.

[4]For instance, if Media-type is video, Media-Suffix includes <frame-id> and <chunk-id>, with the frame itself being either a Key- or I-frame or a Delta- or P-frame.

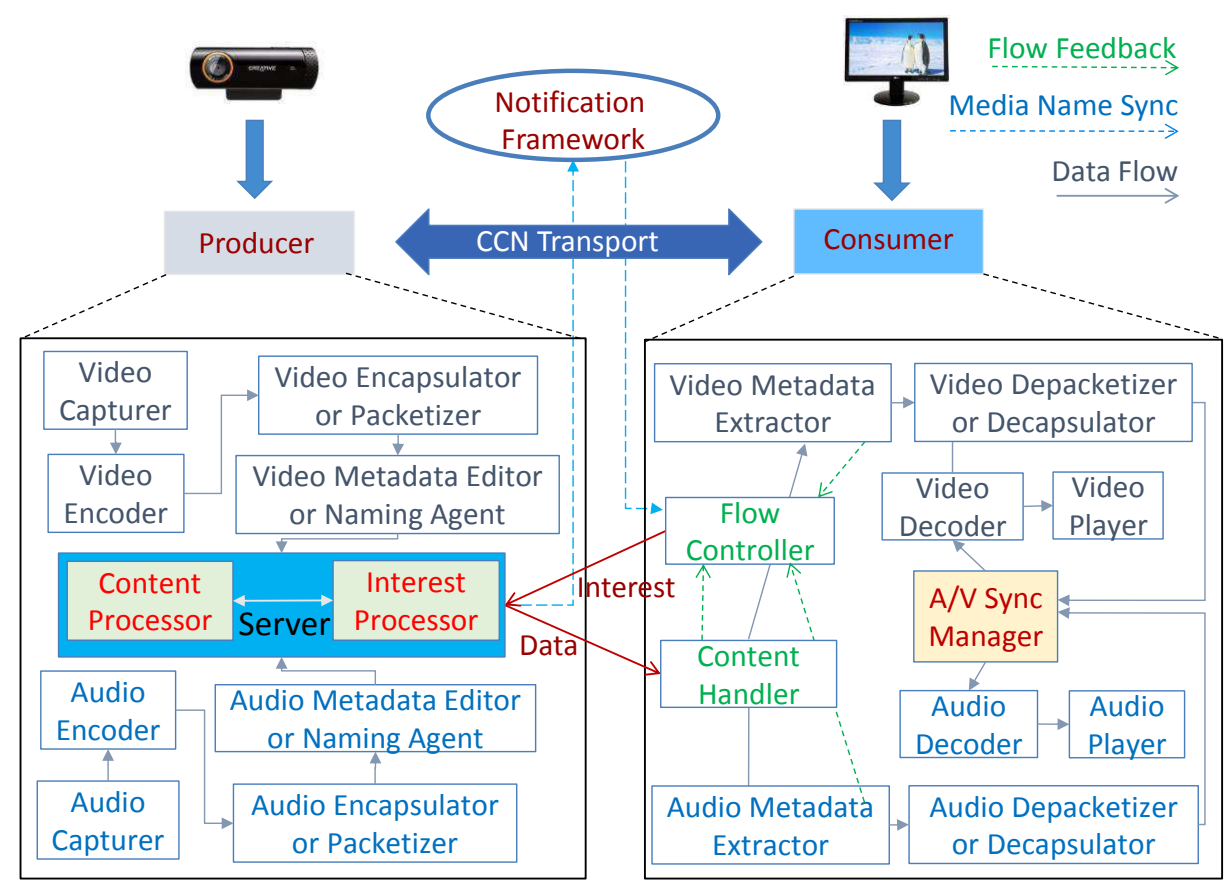

Fig. 3. Producer and Consumer architectures.

Sync-Proxy is also responsible for managing the local notification state of each hosted conference session by keeping a history of updates received from the hosted Sync-Agents and the Sync-Manager. This state can be used for recovery from disruptions. Such a notification state is also managed by the Sync-Agent and the Sync-Manager.

Sync-Manager is responsible for managing the notifications for the hosted conference sessions by relaying fingerprint notifications among the distributed Sync-Proxy instances in a hub-and-spoke manner. Anytime the Sync-Manager receives an update from a Sync-Proxy, the update is pushed to the remote Sync-Proxys. [5]

The frequency at which the notification mechanism operates depends on the media type. In our solution, text-driven notification is for every chat text committed by the participant, whereas real-time content-driven notification is provided at a configurable periodic interval (to manage the associated overhead). In the case of video content, Sync-Client agent sends notifications per multiple group-of-picture (GOP) intervals.[6]

### *C. Conference Data Plane*

As the CCN transport is typically optimized for non-realtime content delivery services, services with stringent latency requirements (*i.e.*, $\leq 150ms$ and $\leq 250ms$ for audio and video services, respectively [17], in addition to the relative audio/video sync requirement of $+45ms$ to

[5]Note that the forwarding mechanism used for the notifications by the CCN Service Controller can be granular to the extent as the notifications are only pushed if there is an interest for them at a given proxy.

[6]Specifically, if the GOP consists of 25 frames, notifications can be sent for every key frame, *i.e*, for every $[25k, 25(k+1))$ frames, where $k \in Z^+$. To reduce signaling overhead, notifications can also be sent for key frames per multiple GOPs.

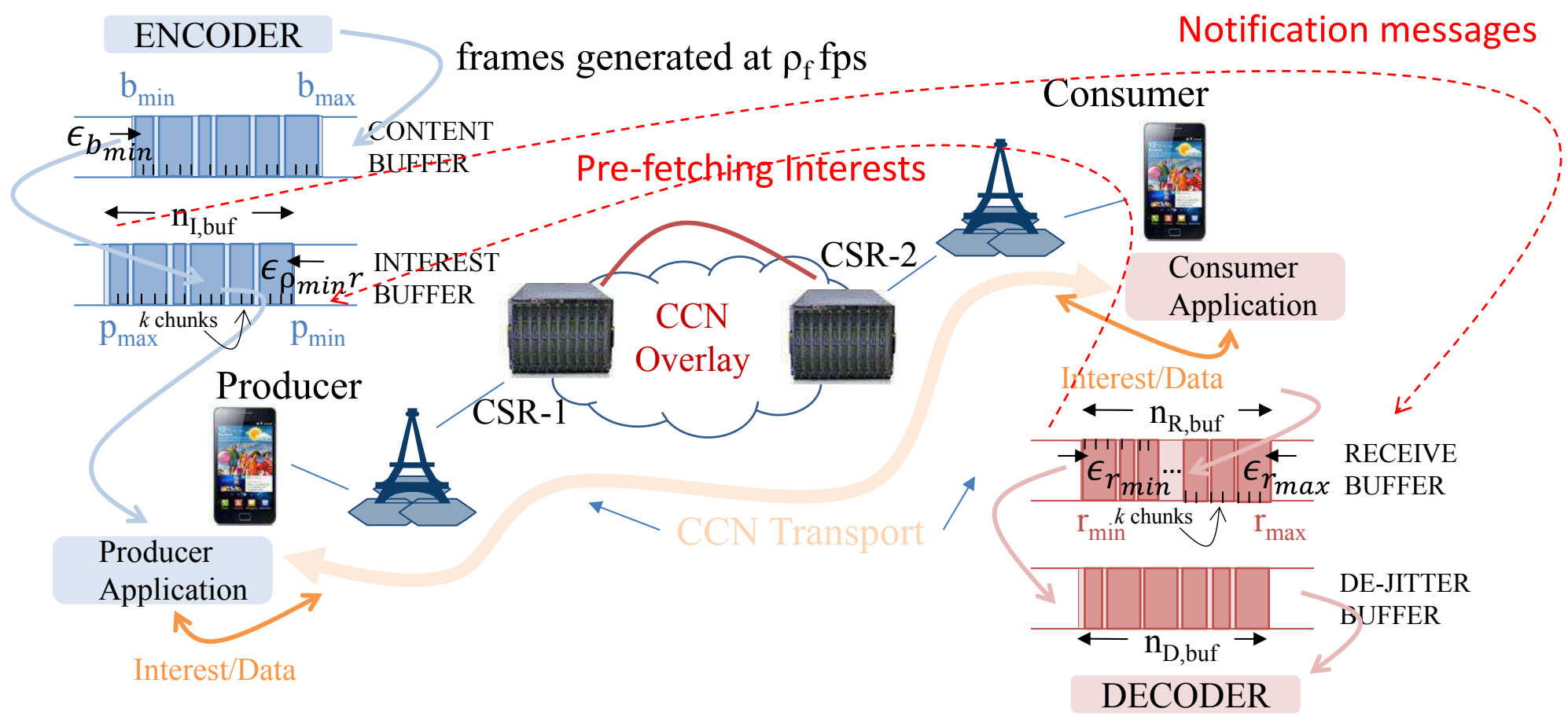

Fig. 4. A high level abstraction for the conferencing framework from the perspectives of the consumer and producer sides.

$-125ms$ [18]) cannot be supported efficiently by applications running on a naive CCN transport.[7]

Considering the strict QoE requirements for the realtime conference service, proposed architecture requires the consumer to pre-fetch the content from the producer, with guidance from the notification framework. To properly handle the pre-fetch content requests, producer implements an application layer buffer to hold the pre-fetching Interests and to enable the content to be pushed to consumers as soon as a content object is generated. We next provide a high level functional design of the producer and consumer nodes with reference to Figure 3.

*1) Producer Design:* At the producer side, `video-encoder`, which receives the live content from the `video-capturer`, provides the application with the encoded key/delta frames. Depending on CCN chunk-size setting, encoded frames are chunked by the `video-data-encapsulator` and appended with the video metadata (*i.e.*, current frame index, frame type, number of chunks) by the `video-metadata-editor`. Next, the `naming-agent` names the content chunks using the procedure described in Section II-A, and publishes them at the `server-module`'s `content-processor` for consumption. This `server-module` also includes an `interest-processor` function, which holds the pre-fetching Interests. Published content is matched against the pre-fetching Interests to satisfy the consumer requests. We discuss in more detail the algorithms executed by the `server-module` in Section III-B. We use a similar pipeline for the audio content, except that, as the payload size per frame typically varies within the $10-50$B range, each frame is encoded separately.[8]

*2) Consumer Design:* At the consumer side, process to display a participant's video is initiated by the `Sync-Agent` through its communication with the `Sync-Proxy`, as part of the notification framework. In doing so, `Sync-Agent` acquires information on the producer's latest video/audio name state, which is sufficient for the consumer to start expressing Interests and sync with the real-time content generated by the producer.

We show the consumer design in Figure 3, where the `flow-controller`, which is the main module connected to the notification framework, is responsible for (*i*) issuing an estimated set of Interests to fetch the real-time content, and (*ii*) assigning a play-out time for the subsequent frames. Here, the main challenge is to correctly determine the variable number of chunks corresponding to a video frame (as video frame sizes change in time, and likely to be from one frame to another). We address this problem using the frame descriptor metadata carried within each video chunk, which is extracted by the `metadata-extractor` and fed back to the `flow-controller`, after receiving the chunks from the `content-handler`.

`flow-controller` uses the periodic notifications it receives from the notification framework to sync with the stream information (so as to fetch content), which also helps with new joins and network disruptions. Note that, during significant network disruptions, which can last in the order of seconds, `flow-controller` starts from the latest content state learned by the notifications, while clearing out the outstanding Interests for the earlier video frames. If the `content-handler` module does not acknowledge the `flow-controller` in regards to receiving a given chunk, `flow-controller` may initiate retransmission to potentially recover from the network cache or directly from the producer. To support the smooth playout, a de-jitter buffer is implemented by the `content-handler`. Proposed design also includes an `A/V Sync-Manager` module (`AVSyncManager`) that uses the timestamp information from the audio/video frames to sync the audio/video playback before handing the content to their respective decoders.

## III. Analytical Study for the Conferencing Framework

In this section, we start by presenting an in-depth analysis of the proposed architecture in Section III-A to understand the important engineering variables and settings at the consumer/ producer sides to meet the end-to-end latency requirements. We show a high level abstraction of the consumer and producer sides for the delivery of the video content in Figure 4.[9] Next, in Sections III-B and III-C, we present the specific algorithms executed by the consumer/producer sides based on the initial analysis. The current analysis is presented at the granularity of frames (as each frame represents the minimum playable unit, even though it may consist of multiple chunks). We present the definitions for the key parameters used in our architecture in Table I.

[7]Note that, latency requirement for the chat service is much less stringent and is in the order of $\leq 1s$.

[8]Note that, another optimization considering the CCN overhead is to carry multiple audio frames within a single content object.

TABLE I
PARAMETER DEFINITIONS

| PARAMETER | DEFINITION |
|---|---|
| $\{b;p;r\}_{\{\min;\max\}}$ | {min; max} frame index in `{content;interest;receive}-buffer` |
| $\delta_{de-jitter}$ | de-jitter buffer length (in time units) |
| $\delta_{\{enc;dec\}}$ | {encoding;decoding} latency |
| $\delta_{pre-fetch}$ | pre-fetching duration |
| $\delta_{rtt}$ | round trip time |
| $\epsilon_i;\epsilon_{ir}$ | {generated;expected} chunks for frame $i$ |
| $\gamma$ | consumer lag factor |
| $\iota(i)$ | key frame index for the $i$th frame's `GOP` |
| $k_{notify}$ | number of `GOP`s in-between notifications |
| $N_\iota$ | index for the latest notification |
| $n_{\{I,R,D\},buf}$ | `{interest;receive;de-jitter}-buffer` size |
| $\rho_f;t_f$ | frame {rate;duration} |
| $\sigma$ | `GOP` size |
| $\tau_i$ | play-out time for the $i$th frame |
| $\tau_{notify}$ | notification interval |
| $T^{\iota(i)}_{notify}$ | notification receive time for $\iota(i)$ |
| $\{\theta;n_\theta\}$ | fetch-ahead {time;gap} |

### A. System Analysis

For the current analysis, we assume that the consumer initiates a session by requesting video content $\theta$ms into the future with respect to the latest received notification $N_\iota$ with $\theta$ representing the *fetch-ahead* duration. We assume the producer to generate video at a rate of $\rho_f$ frames per second (*fps*) with each frame assumed to encode $t_f$ms duration of content.

† **Producer-side components:** As shown in Figure 4, producer manages two buffers (which are managed by the producer's `server-module`): (*i*) `content-buffer`, which holds the content generated by the `video-encoder` and published as `CCN` content-objects, and (*ii*) `interest-buffer`, which holds the pre-fetching Interests sent by the consumer. At the `content-buffer`, we identify the min (or max) frame index window with $b_{\min}$ (or $b_{\max}$), with each frame $i$ carrying $\epsilon_i$ chunks.[10] At the `interest-buffer`, we identify the min (or max) frame index window for the requested Interests with $p_{\min}$ (or $p_{\max}$), where the number of pre-fetched chunks for a frame $i$ is denoted with $\epsilon_{ir}$.

The maximum size–in number of frames–for the pre-fetching `interest-buffer` window is equal to $n_{I,buf}$.[11] Producer issues notifications that coincide with the generation of a key frame within a given `GOP`, for which the size–in time units–is given as $\sigma$. We represent the notification interval with $\tau_{notify}$.[12] ‡

† **Consumer-side components:** Similar to producer case, consumer manages two buffers as well: (*i*) `receiver-buffer`, which is of length $n_{R,buf}$ and used to store pre-fetching Interests, and (*ii*) `de-jitter buffer`, which contains only the completed frames and is of length $n_{D,buf}$ and used to smooth video play-out.

The min (or max) frame index at the `receive-buffer` for requested frames is identified as $r_{\min}$ (or $r_{\max}$). The consumer algorithm may fetch more or less depending on the number of chunks estimated by the consumer's fetch algorithm. If the consumer under-fetches, the difference (*i.e.*, $\epsilon_i$-$\epsilon_{ir}$) is inferred from the metadata within the received content-objects, and the missing chunks are subsequently requested from the producer. ‡

We next analyze the impact of the producer/consumer side components (and the chosen parameters) on the end-to-end video latency. We assume that $\delta_{e2e}$ represents the expected end-to-end video latency and $\delta_{rtt}$ represents the round-trip-time (`RTT`) between the producer and the consumer.

† We can determine the one-way latency for video delivery (under the pre-fetching assumption) after a content has been generated at the producer, $\Delta^{\rightarrow}$, as follows:

$$\Delta^{\rightarrow} = \delta^{\rightarrow} + \delta_{de-jitter} + \delta_{codec} \tag{1}$$

where $\delta^{\rightarrow}$ represents the time to transfer the chunks of a frame from producer to consumer, $\delta_{de-jitter}$ represents the de-jitter buffer delay (*i.e.*, $\delta_{de-jitter} = n_{D,buf} \times t_f$ ), and $\delta_{codec}$ represents the encoding/decoding delay, which we can express as the sum of encoding and decoding latency components $\delta_{codec} = (\delta_{enc} + \delta_{dec})$.

We can express the latency requirement corresponding to video content as follows:

$$\delta_{rtt} \times \left(\frac{1}{2} + \lceil \texttt{sgn}(\epsilon_i - \epsilon_{ir})\rceil^{+}\right) + \delta_{de-jitter} + \delta_{codec} \leq \delta_{e2e} \tag{2}$$

where the first component represents the time to transfer all the chunks of a frame, including the missing chunks (assuming the missing chunks are requested in batch), `sgn` represents the sign function[13], and $\lceil x\rceil^{+}$ represents $\max(x, 0)$. Here, the assumption is that if $\epsilon_i > \epsilon_{ir}$, then a batch request of Interests are made to retrieve the remaining chunks of the $i$th frame, thereby incurring a delay of *rtt* duration. ‡

† At the consumer side, we can calculate for frame $k$ the play-out time, $\tau_k$, with respect to the expected receive time of its corresponding notification, $T^{\iota(k)}_{notify}$, as follows:[14]

$$\tau_k = \left[\delta_{e2e} - \left(\delta_{rtt} + \delta_{dec} + D_j\right)\right] + \theta + T^{\iota(k)}_{notify} \tag{3}$$

[9]A similar analysis can be applied on the audio content.

[10]Number of chunks depends on the engineered `CCN` chunk size of $C_s$.

[11]Note that, $n_{I,buf}/\rho_f$ gives the duration for the pre-fetched content, in time units.

[12]We can express $\tau_{notify}$–in number of `GOP` periods–using $\tau_{notify} = k_{notify} \times \sigma$, where $k \in Z^{+}$.

[13]`sgn(x)`=:{$-1$ if $x < 0$},{0 if $x = 0$},{1 if $x > 0$}.

[14]Note that, the notification time represents the consumer side system time (*i.e.*, clock value), at the time the consumer receives the notification corresponding to a `GOP` or multiple `GOP`s.

if $k = \iota(k) + \frac{\theta}{t_f}$ (where $\iota(k)$ represents the key frame index corresponding to the received notification) and after the consumer has just joined the conference and received the first notification; and

$$\tau_k = \tau_j + t_f \tag{4}$$

if $k = j + 1$, and after the consumer has realized that it is out-of-sync with the producer, where the out-of-sync threshold is based on the difference between the index of the latest received notification $N_\iota$ (*i.e.*, $\iota(k)$) and $r_{\min}$. Here, the subsequent frame's play-out time is determined by adding the frame duration to the previous frame's play-out time. ‡

† After the play-out times are assigned to each frame, we can express the requirement corresponding to the play-out deadline as follows:

$$\delta_{play-out} \geq \delta_{pre-fetch} + \delta^{\rightarrow} + T_{notify} \tag{5}$$

where, for the $k$th frame, $\delta_{play-out}$ equals $\tau_k$, $T_{notify}$ equals $T^{\iota(k)}_{notify}$,

$$\delta^{\rightarrow} = \delta_{rtt} \times \left( \frac{1}{2} + \lceil \texttt{sgn}(\epsilon_k - \epsilon_{kr}) \rceil^{+} \right) \tag{6}$$

and

$$\delta_{pre-fetch} = \theta + \frac{n_{I,buf}}{\rho_f} \tag{7}$$

‡

Based on the above understanding of the system and its components, we next discuss the consumer and producer side algorithms and present their respective pseudo codes, in the context of video content.[15] For the presented algorithms, the appropriate parameter settings to handle certain events (*i.e.*, network congestion or host connectivity issues) have to be determined using simulation studies and further refined using prototype implementations.

## B. Producer-side Algorithms

At the producer side, two main algorithms are implemented (which are simpler in logic compared to the algorithms run by the consumer): (*i*) *content publish* algorithm, which is executed by the `content-processor` module and for which the pseudo code is given by `Algorithm 1`, and (*ii*) *interest processing* algorithm, which is executed by the `interest-processor` module and for which the pseudo code is given by `Algorithm 2`.

*1) Content Publish Algorithm:* As the producer receives the key/delta frames from the encoder, received frames are first chunked based on the pre-defined `CCN` chunk size, and then the associated metadata is inserted within each chunk (`Steps 2-6`). If the notification interval $\tau_{notify}$ divides the frame index $i$ (*i.e.*, $\texttt{mod}(i \times t_f, \tau_{notify}) = 0$, where `mod` represents the modulo operation), then the frame is considered a key frame, and a notification is issued for this key frame (`Steps 8-10`). Once a content-object is produced, its name is compared with that of the pre-fetching Interests stored within the `interest-buffer`, and if there is a match, then the content-object is returned to the requesting consumers (`Steps 11-13`). The `interest-buffer` and `content-buffer` variables ($p_{\min}$ and $b_{\min}$) are then incremented (by one or multiple frame indexes, if required) depending on whether the current content-object can satisfy all the content-object chunks corresponding to the $p_{\min}$ frame (`Steps 14-15`).

*2) Interest Processing Algorithm:* This algorithm is executed whenever an Interest arrives at the producer (`Step 1`). These Interests are either pre-fetching Interests (`Steps 5-8`), or Interests for frames that were under-fetched (`Steps 2-4`). As the incoming Interests fill a frame, $p_{\max}$ is incremented (`Step 6-8`). The algorithm then checks the difference of $\big(p_{\max} - b_{\max}\big)$, which is expected to be close to the fetch-ahead gap given by $n_\theta = \big(\theta \times \rho_f\big)$. If the difference is less than a factor of $n_\theta$ (*i.e.*, $p_{\max} - b_{\max} < \omega \times n_\theta$, where $\omega \in (0, 1)$), then the algorithm infers that the consumer is under requesting, which requires the notification frequency to be increased. This is done by reducing $\tau_{notify}$ (`Steps 10-14`). Next, the algorithm checks the fairness status, by determining whether the consumers are being unfair while pre-fetching content. For this purpose, the algorithm checks if the difference $\big(p_{\max} - b_{\max}\big)$ is greater than $\big(\xi \times n_\theta\big)$, where $\xi > 1$, in which case the Interest that satisfies the condition is dropped (`Steps 15-17`).

**Algorithm 1** Producer Content Publish Algorithm

**Require:** $\rho_f$, $\tau_{notify}$, $l_{chunk}$

1: Set `frame_index` $i = 0$, and `content-buffer` index $b_{\min} = 1$
2: **for** each encoded frame received from `encoder` **do**
3: $b_{\max} = b_{\max} + 1$;
4: Set `frame_size` $l_{frame} = sizeof(frame(i))$ Bytes;
5: $\epsilon_i = l_{frame}/l_{chunk}$ chunks;
6: Insert metadata into each chunk
7: *`Publish chunks`* as named content-objects in the `content-buffer`;
8: **if** $(i \times t_f) \pmod{\tau_{notify}} = 0$ **then**
9: *`Issue notification`* for current key-frame with index $i$;
10: **end if**
11: *`Perform`* chunk-to-(pre-fetching)*`Interest matching`*;
12: **if** names match for chunk and pre-fetching Interest **then**
13: *`Forward`* the matching *`content-object`* towards consumer;
14: **if** Forwarded content-object fills $p_{\min}$ frame **then**
15: Set $p_{\min}$ to min frame index with non-fulfilled Interests, and increment $b_{\min}$ accordingly;
16: **end if**
17: **end if**
18: $i = i + 1$;
19: **end for**

## C. Consumer-side Algorithms

At the consumer side, three main algorithms are executed: (*i*) *notification handling* algorithm, which is executed by the `flow-controller` module and for which the pseudo code is given by `Algorithm 3`, (*ii*) *pre-fetching* algorithm, which is also executed by the `flow-controller` module and for which the pseudo code is given by

[15]Note that, algorithms for the audio content follow similar logic except that the shorter frame sizes for the audio content afford not using chunking.

**Algorithm 2** Producer Interest Processing Algorithm

**Require:** $n_\theta, k_{notify}, \sigma, \omega, \xi$, where $\omega \in (0,1)$, $\xi > 1$

**for** every admitted Interest to `interest-buffer` **do**
  **if** Interest has a match in `content-buffer` **then**
    *`Forward`* matching *`content-object`* to consumer;
    Set $p_{\min}$ to min frame index with non-fulfilled Interests, and increment $b_{\min}$ accordingly; {Note: Non-fulfilled Interests correspond to under-fetched frames.}
  **else if** Interest fills a frame in the `interest-buffer` **then**
    **if** frame index $> p_{\max}$ **then**
      $p_{\max} =$ `frame_index`;
    **end if**
  **end if**
  **if** $p_{\max} - b_{\max} < \omega \times n_\theta$ **then**
    **if** $k_{notify} > 1$ **then**
      $\tau_{notify} = (k_{notify} - 1) \times \sigma$; {Note: As consumers are not expressing enough Interests, notification rate is increased.}
    **end if**
  **end if**
  **if** $p_{\max} - b_{\max} > \xi \times n_\theta$ **then**
    *`Remove received Interest`*; {Note: As consumers are behaving unfairly, Interests with indexes above $\xi \times n_{I,buf}$.}
  **end if**
**end for**

**Algorithm 3** Consumer Notification Processing Algorithm

**Require:** $\iota$, $n_\theta$, $n_{R,buf}$, $\beta$, $\gamma$, where $\{\beta, \gamma\} \in (0,1)$

1: `Initialization`: participant joins the conference session.
2: **if** Notification from `Sync-Proxy` for frame index $i$ is the 1st **then**
3:   $r_{\min} = \iota(i) + n_\theta$;
4:   $r_{\max} = r_{\min} + n_{R,buf}$;
5:   *`Start`* `Algorithm` 4;
6: **else**
7:   **for** every new notification $N_\iota$ with frame index $i$ **do**
8:     **if** $\iota(i) > r_{\min} + n_{R,buf}$ **then**
9:       $r_{\min} = \iota(i) + n_\theta$;
10:       $r_{\max} = r_{\min} + n_{R,buf}$; {Note: Possible network failure, reset `pre-fetch buffer`, and start `Algorithm` 4.}
11:     **else if** $\iota(i) > r_{\min} + \gamma \times n_{R,buf}$ **then**
12:       $r_{\min} = \iota(i) + \beta \times n_\theta$;
13:       $r_{\max} = r_{\min} + n_{R,buf}$; {Note: As consumer lags behind, we discard a few earlier frames and advance the pre-fetch window, and start `Algorithm` 4.}
14:     **end if**
15:   **end for**
16: **end if**

`Algorithm 4`, and (*iii*) *content-object processing* algorithm, which is executed by the `content-handler` module and for which the pseudo code is given by `Algorithm 5`.

*1) Notification handling Algorithm:* This algorithm is responsible for managing the received producer notifications, and implements the steps corresponding to the three possible cases. The first case corresponds to initialization phase, whereas the last two cases take place during regular operation.

The *first case*, which is explained through `Steps 1-5`, happens when the consumer joins a conference session and receives the first notification, in which case, $r_{\min}$ is set to the index of the frame that is $\theta$ms ahead of the current time and $r_{\max}$ is set to the size of the receive-buffer, while the *pre-fetching* algorithm is bootstrapped. After that, anytime a notification is received, the algorithm compares the current notification frame index with that of the oldest unsatisfied frame request within the receive-buffer, and during normal operation of the protocol, expected distance between the two should be small.

The *second case*, as explained through `Steps 6-10`, corresponds to consumer going out-of-sync with the producer, in which case the difference exceeds the pre-defined threshold $\epsilon_{sync} \times n_{R,buf}$, where $\epsilon_{sync} \in (0,1)$, and requires the consumer to conservatively move the receive-buffer window by a factor of $\varepsilon_{sync} \times \beta$, where $\varepsilon_{sync} \in (0,1)$.

The *third case*, as explained through `Steps 11-14`, is an extreme version of the second case (*e.g*, consumer experiencing a disconnect), in which case, the difference exceeds the receive-buffer length of $n_{R,buf}$, requiring both $r_{\min}$ and $r_{\max}$ to be reset similar to a situation when the consumer receives the first notification.

*2) Pre-fetching Algorithm:* This algorithm, which is bootstrapped by the notification handling algorithm, initiates, as explained through `Steps 1-5`, by expressing Interests (in batch mode) for frames with indexes within the range $i \in (r_{\min}, r_{\max})$, while setting their respective play-out times $\tau_i$. After the initial startup phase concludes, Interests are issued at intervals of $1/\rho_f$, with the play-out time set relative to the previous frame `Steps 7-10`. For the given Interest issuance rate, this algorithm also checks if $r_{\min}$ and $r_{\max}$ have been reset by the notification algorithm (corresponding to failure scenarios), in which case, receiver-buffer window is progressed by discarding Interests with indexes smaller than the current $r_{\min}$ and expressing new Interests for frames up to the current $r_{\max}$ (`Steps 11-16`). If the new $r_{\min}$ is outside the current receive-buffer window (*i.e.*, $r_{\min} > r^*_{\max}$), then the algorithm starts afresh (`Steps 17-19`).

*3) Content-object processing Algorithm:* This algorithm initiates after a content-object (*i.e.*, $pkt_{i,j}$, chunk $j$ of frame $i$) is received by the consumer `Step 1`. First, the algorithm checks if the received frame $i$ is within the receive-buffer window (`Steps 2-3`). If so, metadata from the content chunk is extracted to determine whether or not sufficient number of chunks have been requested from the producer for the given frame `Step 4`. If the number of requested chunks is less than the frame length (in number of chunks), then the pre-fetching algorithm is notified to request for the missing chunks (`Steps 5-7`). On the other hand, if more chunks were pre-fetched by the consumer, then the Interests corresponding to these chunks are cancelled in `CCN` to prevent retransmission attempts (`Steps 8-9`).

If the received chunk fills the $r_{\min}$th frame, then the index for $r_{\min}$ is incremented to the least incomplete frame index (`Steps 11-13`). If the frame index for the received content-object is outside the current receive-buffer (suggesting an earlier connection failure), then the window was reset by the notification, also requiring the received content-object to be dropped (`Steps 15-18`).

**Algorithm 4** Consumer Pre-fetching Algorithm

**Require:** $t_f$, $\iota$, $\delta_{e2e}$, $D_j$, $\delta_{dec}$, $\theta$ {Note: $r_{\min}$ and $r_{\max}$ are shared variables with `Algorithms` 3 and 5.}
1: `Read` $r_{\min}$, $r_{\max}$, and $T^{\iota}_{notify}$;
2: **for** $k = r_{min}$ to $r_{max}$ **do**
3: **if** $k = r_{\min}$ **then**
4: $\tau_k = T^{\iota(k)}_{notify} + \theta + \delta_{e2e} - \big(D_j + \delta_{dec} + \delta_{rtt}/2\big)$;
5: **else**
6: $\tau_k = \tau_{k-1} + t_f$;
7: **end if**
8: *`Request content-objects`* for frame index $k$ by expressing Interests for $\epsilon_{kr}$ chunks;
9: **end for**
10: **for** every $t_f$ time unit **do**
11: **if** $k = r_{\max}$ **then**
12: {Note: To check if $r_{max}$ was changed by `Algorithm 3`.}
13: $k = k + 1$; $r_{\max} = k$; $\tau_k = \tau_{k-1} + t_f$;
14: *`Express Interests`* for frame $k$ for its $\epsilon_{ir}$ chunks;
15: **else if** $k < r_{\max}$ **then**
16: {Note: `Algorithm 3` made a jump.}
17: **if** $r^*_{\min} < r_{\min} < r^*_{\max}$ **then**
18: {Note: As consumer falls behind producer, pre-fetch window is advanced while discarding out-of-window Interests.}
19: *`Discard Interests`* for frames $r^*_{\min}$ to $r_{\min}$;
20: *`Generate Interests`* for frames $r^*_{\max}$ to $r_{\max}$;
21: **else if** $r_{\min} > r^*_{\max}$ **then**
22: goto `Step 1`
23: **end if**
24: **end if**
25: $r^*_{\min} = r_{\min}$; $r^*_{\max} = r_{\max}$;
26: **end for**

**Algorithm 5** Consumer Content-object Processing Algorithm

1: **for** each frame chunk $pkt_{i,j}$ received from `CCN` **do**
2: *`Read`* $r_{\min}$ and $r_{\max}$;
3: **if** $r_{\min} \leq i \leq r_{\max}$ **then**
4: *`Extract metadata`* $\epsilon_i$ from $chunk_i$;
5: **if** $\epsilon_i > \epsilon_{ir}$ **then**
6: *`Notify`* `flow-controller` module for missing chunks;
7: *`Express requests`* for the missing chunks through `flow-controller`;
8: **else if** $\epsilon_i < \epsilon_{ir}$ **then**
9: *`Notify`* `flow-controller` module on over-fetching; {*`Cancel retransmission`* for outstanding Interests through `flow-controller`;}
10: **else**
11: **if** $pkt_{i,j}$ fills the $r_{\min}$th frame **then**
12: *`Notify`* `flow-controller` about completion of the $r_{\min}$th frame;
13: **end if**
14: **end if**
15: **else if** $r_{\min} > i$ **then**
16: {Note: `Algorithm 3` made a jump.}
17: *`Discard`* $pkt_{i,j}$;
18: **end if**
19: **end for**

## IV. Performance Evaluation

In this section, we evaluate the performance of the proposed architecture. We start by explaining the framework used for our analysis and then present the experimental results for the considered performance metrics.

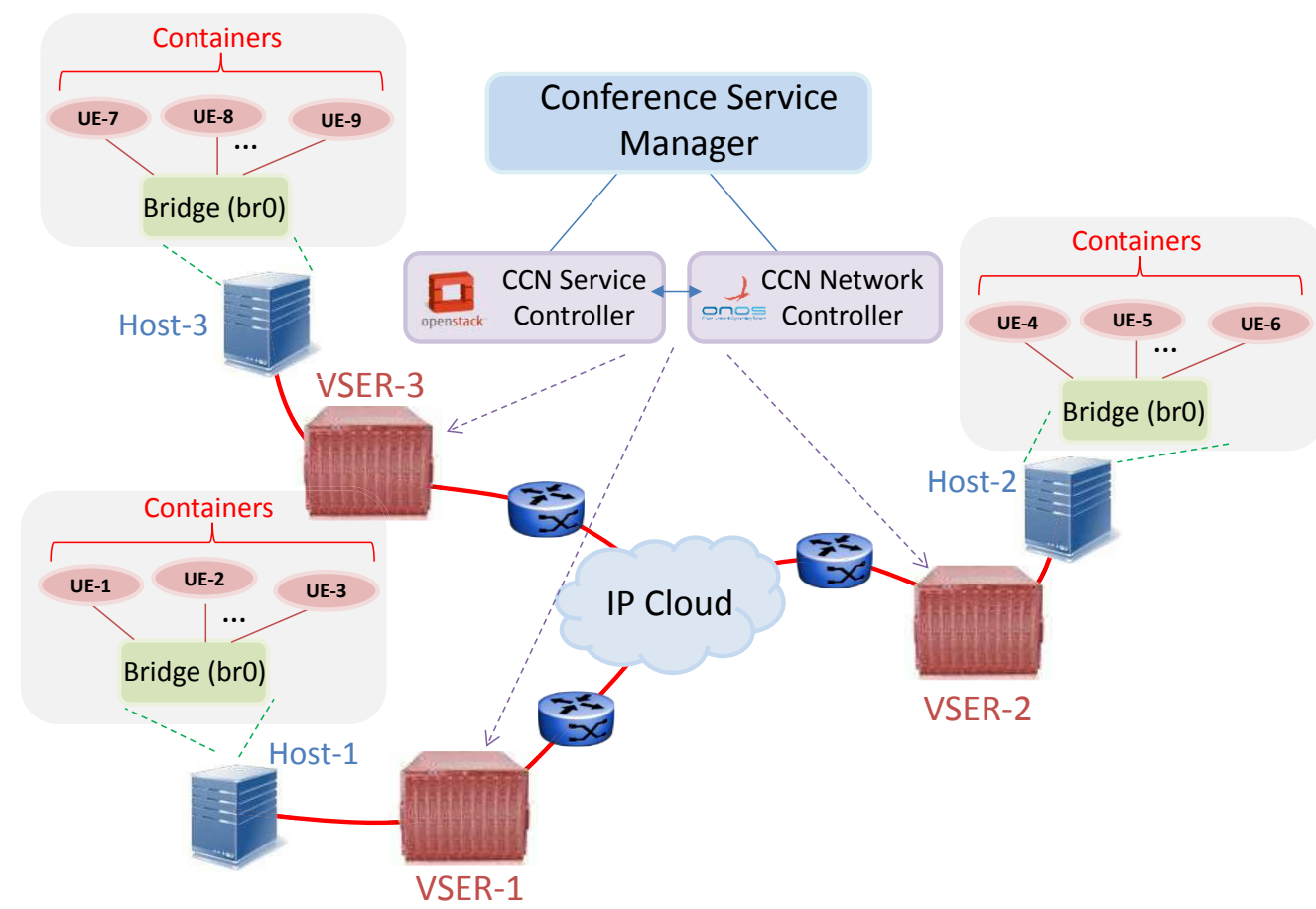

Fig. 5. Emulation test bed topology.

### A. Evaluation Framework

We realized the multimedia conferencing solution over a `CCN/NDN`-based service platform, consisting of programmable `CSR`s, which we also refer as the *Virtual Service Edge Routers* (`VSER`s), [19] that are controlled by application controllers interfacing with `OpenStack` [14] and `ONOS` [15] frameworks. Here, `OpenStack` manages the placement of service-`VM`s such as `Sync-Proxy` and `Sync-Manager` within `VSER`s, while `ONOS` handling dynamic service routing using a `Conference Network Controller` (realized as an application in `ONOS`) based on specific service events from the service-`VM`s in the `VSER`s as shown in Figure 4.

The prototype for the multi-party conferencing architecture (which is shown in part of Figure 5) was demonstrated with 5 participants and the results were published in [19]. In this paper, to study the system at scale, we instead used an emulation platform to emulate the media sources for the consumers and the producers. For the emulation platform, we used `NDN` as the code base, which is different from our original design [19] where we used `CCNx` [20]. The emulation test-bed consists of 5 host machines and 3 service nodes with `VSER`s overlaid over the `IP` network, which is conceptually shown in Figure 5 with 3 host machines. We used a single `IP` router to represent the `IP` cloud. To emulate the `UE`s, we utilized Linux Containers installed on the 5 host servers, and distributed the `UE`s in a round robin manner to individual containers on the host machines. In the current study, we varied the number of participants from 1 to 15. For example, for the 3-`UE` case, we utilized 2 host machines with one running 2-`UE` containers and the other running a single `UE` container; and for the 15-`UE` case, we utilized all the host machines, with each running 3 `UE` containers.[16] Random link latencies (of 40 ∓ 10ms) and bandwith limits are introduced at the container level to emulate link quality variations for the end-to-end connections, which, for instance, allow for non-symmetric latencies. We used Intel Skylake $i7-6700$ 3.4Ghz 4-core (8-thread) machines for the UE-hosting servers and the IP routers, and 2-to-4 core machines for the VSERs (specifically, VSER-1 is implemented over an Intel Skylake $i7-6700$ 3.4Ghz 4-core (8-thread) machine, while the other VSERs were implemented on Intel Haswell $i7-4600U$ $2.1GHz$ 2-core (4-thread) machines).

[16]Specifically, 6-client case assigns 2-`UE` containers to one machine, and 1-`UE` container each to the rest; 9-client case assigns 2-`UE` containers on four machines and 1-`UE` container to one; and 12-client case assigns 3-`UE` containers to two machines and 2-`UE` containers to 3 machines. To represent the lower-bound on the perceived performance, for each case, client side measurements are reported at the host machine that runs the highest number of `UE` containers.

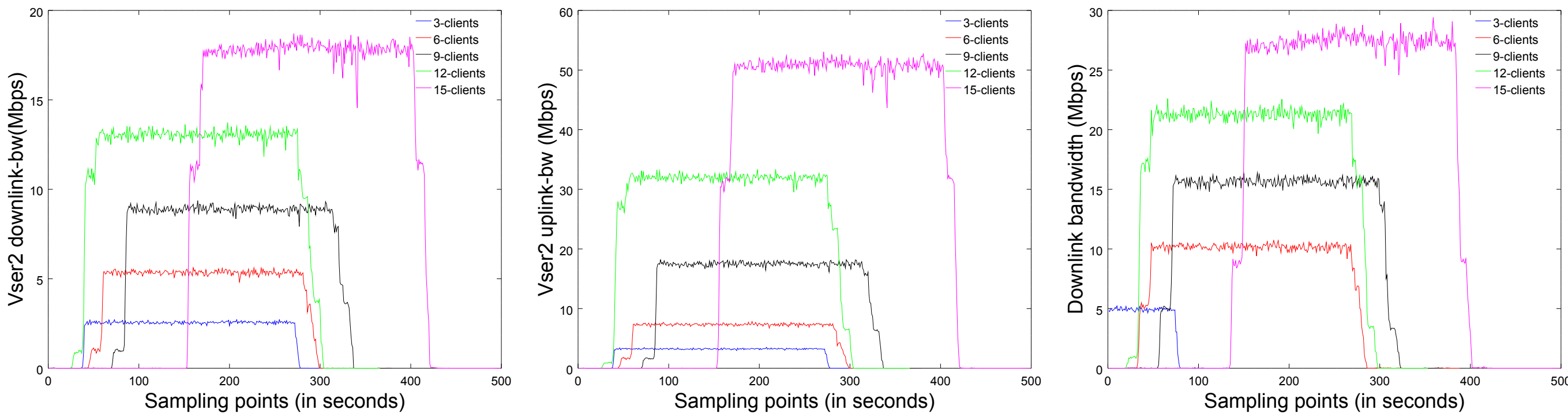

(a) VSER downlink bandwidth utilization (b) VSER uplink bandwidth utilization (c) IP router bandwidth utilization

Fig. 6. Bandwidth utilization at different architectural components.

To emulate the video content generated by the UEs, we used a video traffic model that was derived from our earlier prototype (based on the statistics for the Key/Delta frame sizes captured over the course of a 10min conferencing session). To emulate the audio content, which was modeled after G.729 codec [21] generating constant bit rate (CBR) traffic, we used a CBR traffic of 30Kbps. The algorithms discussed in Sections III-B and III-C introduced several tuneable parameters for handling notifications, requests and content objects. We empirically derived the suggested parameters for our setting (for which the analysis is omitted due to space considerations) and show their values in Table II.[17]

TABLE II
SELECTED PARAMETER VALUES

| Parameter | Value |
|---|---|
| $\rho_f$;$t_f$ | 25fps;40ms |
| $\sigma$;$\tau_{notify}$ | 25fms;25fms |
| $\theta$ | 1s |
| $n_{\{I;R;D\},buf}$ | 100fms;100fms;40ms |
| $l_{chunk}$ | 3000Bytes |
| $\epsilon_{ir}$\{Key;Delta\} | \{5;1\} chunks |
| $\gamma$ | 0.5 |

In current evaluations, our primary goal is to demonstrate the scalability and reliability of the multi-party conferencing architecture, under various scenarios. From the UE's perspective, we use end-to-end application level latency, effective throughput, and computing utilization as the primary performance metrics. As stated earlier, end-to-end latency performance benchmark for audio/video content is set to $\leq 150$ms/$\leq 250$ms respectively [17]. From the network perspective, we will use bandwidth utilization and computing efficiency as the main performance metrics at the VSER nodes and the IP router.

## B. Bandwidth Utilization

We show the bandwidth utilization as measured by the IP traffic at different network components, such as VSER node and the IP router in Fig. 6.[18] Here downlink bandwidth refers to the incoming traffic at the given host, whereas uplink traffic refers to the outgoing traffic.

As the contents generated by a client is first transmitted to its servicing VSER node, and from there unicast transmitted to the other VSER nodes servicing active clients, we can approximate the downlink bandwidth utilization at the $i$th VSER node, $W^{\texttt{dl}}_{\texttt{VSER},i}$, using the following equation:

$$W^{\texttt{dl}}_{\texttt{VSER},i} = w^{(\texttt{I})}_{\texttt{UE}} \times \kappa_{\texttt{VSER},i} \times (\kappa_u + n_{\texttt{VSER}} - 2) + w^{(\texttt{D})}_{\texttt{UE}} \times \kappa_{\texttt{VSER},i} \quad (8)$$

where $n_{\texttt{VSER}}$ represents the number of VSER nodes, $\kappa_{\texttt{VSER},i}$ represents the number of clients serviced by $\texttt{VSER}_\texttt{i}$, $\kappa_u$ represents the total number of clients serviced by all VSERs (for the given session), $w^{(\texttt{I})}_{\texttt{UE}}$ represents the average generated interest stream rate towards a client, and $w^{(\texttt{D})}_{\texttt{UE}}$ represents the average per-client generated data stream rate. Here, for the sake of simplicity, we assume the clients to generate similar traffic streams. Our analysis can easily be extended to cover the case for different stream rates. Similarly, we can express the uplink bandwidth utilization at the given VSER node as follows:

$$W^{\texttt{ul}}_{\texttt{VSER},i} = \; = w^{(\texttt{D})}_{\texttt{UE}} \times \kappa_{\texttt{VSER},i} \times (\kappa_u + n_{\texttt{VSER}} - 2) + w^{(\texttt{I})}_{\texttt{UE}} \times \kappa_{\texttt{VSER},i} \quad (9)$$

The results closely match with our expectations, as we increase the number of clients participating in the multi-party conferencing session, pointing to a linear increase in bandwidth requirements.[19] We also observed that the control message overhead (*i.e.*, notifications) is less than 2% of the overall overhead for all the considered scenarios, limiting its impact on the overall performance.

[17]Note that, as de-jitter buffer is not part of the current implementation, we account for its impact using a static delay, which was integrated to our end-to-end latency calculations. Additionally, current implementation lacks dynamic notification adaptation and producer side check for unfair Interest expression by a consumer.

[18]As the clients are evenly distributed to both the host machines and the VSER nodes, all the VSER nodes experience similar traffic patterns, hence, for brevity, we are only reporting the link utilization results at one of them.

[19]Specifically, underestimating the observed results by less than 2% due to unaccounted overhead within calculations, such as retransmissions, notifications, and protocol overheads.

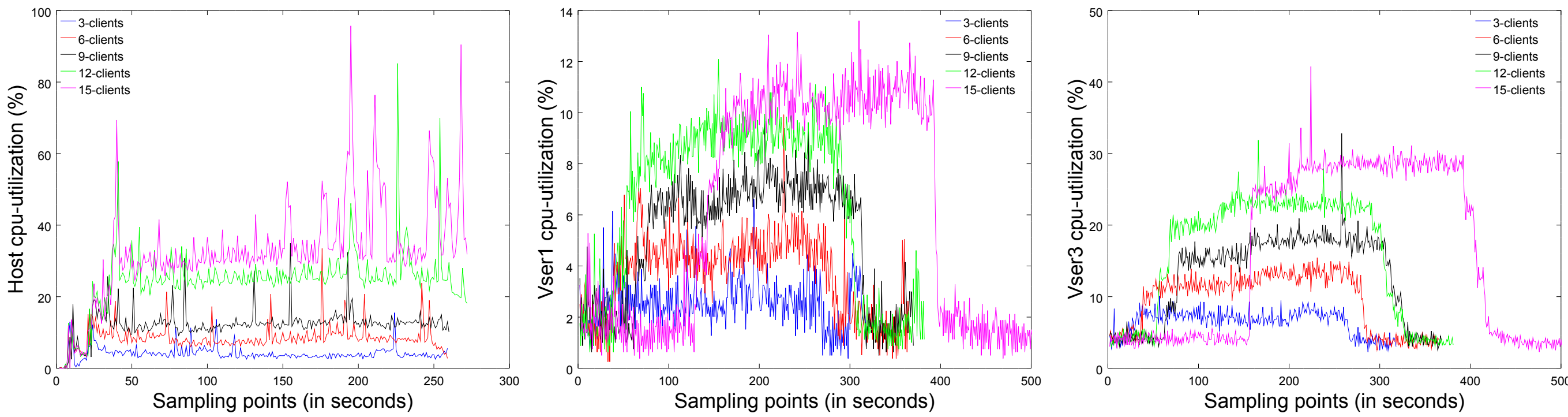

(a) CPU use at the host server (b) CPU use at VSER-1 (c) CPU use at VSER-3

Fig. 7. CPU utilization at different architectural components.

### *C. CPU Performance*

We illustrate the CPU utilization performance in Figure 7, which shows the average CPU utilization across all cores at the considered nodes. We observe that at both the client hosting machine (results are from the first machine, which gets assigned the most client containers) and the VSER nodes, CPU utilization approximates a linear increase in the number of participants even though the architecture utilizes an IP-based overlay (for the selected chunk size, for instance, experiencing IP (de-)fragmentation at the CCN hosting servers such as host machine and VSER nodes). VSER nodes observe similar normalized performances, with the main difference being attributed to the different number of cores utilized for their respective implementations (*i.e.*, VSER-1 node having 8-cores vs. VSER-3 (or VSER-2) having 4-cores).

At the client host machines, we observed around 35% CPU utilization with 15 clients (with each host machine carrying 3-UE containers). Note that, the nonlinear performance increase, going from 9 clients to 12 clients, can be explained with the increased number of both client containers and the number of total participants (*i.e.*, going from 2 to 3 client containers, and from 9 to 12 participants). If we proportionate these values to number of clients hosted by the server node (which would be the more practical scenario), the results suggest approximately 12% CPU utilization per client with 15 participants acting as both Producer and Consumer. Based on these results, for a setup like ours with 5-host machines, a client with the compute specs used in our design shall be able to support more than 100 participants (assuming the bandwidth requirements are also met to deliver all those streams).[20]

If we compare the results at the host machines and VSER-1, both of which share 8-cores, we can draw similar conclusions, with capability to support more than 100 participants. To illustrate the capability for the architecture to support more nodes (while isolating the impact of one consumer on another due to hosting their containers on the same machine), we tested our architecture using a single consumer and multiple producers, where we varied the number of producers from 18 to 45. Figure 8(a) illustrates the CPU utilization at the host machine as we increase the number of producers. We observe less than 30% CPU utilization with 45 producers, with little loss in performance as we increase the number of producers.

### *D. Scalability Performance*

In Figure 8(b) we illustrate the end-to-end latency performance for both the audio and video streams from the point of view of a single consumer as we increase the number of participants, with each participant acting as both a consumer and a producer. In our experiment, a new participant joins the conferencing session at regular intervals, with each consumer executing the pre-fetching process soon after receiving the first notification from its sync-proxy. As the Interests are pre-fetched, latency here represents the one way delay from the producer with respect to the audio or video frame generation time up until when the frame is ready for decoding at the consumer side. For both audio and video content, as we increase the number of participants from 3 to 15, the latency values for the majority of frames stay below the critical threshold based on our definition of service quality, *i.e.*, less than 150ms for the audio stream and 250ms for the video stream. We observe that the 3 participant case for the delivery of video streams performs poorly compared to delivery of audio streams or with respect to the other multi-participant conferencing scenarios. This is because the video frames involve more chunks than fetched by the pre-fetch logic resulting in a second phase of data retrieval. Multi-participant situations take advantage of the caching/aggregation feature to achieve better performance here.

We also observe that as we increase the number of participants the results worsen at a higher rate, specifically going from 12 participant to 15 participant case. We observe a different level of worsening in performance (in terms of frame loss) for the single-consumer multi-producer case, for which the results are shown in Figure 8(c). The performance loss can be explained by (*i*)

[20] Note that, if we divide the CPU utilization with $\kappa_{vser} \times (\kappa_{vser} \times n_{vser} - 1)$, we observe almost the same average CPU utilization value, which is $\approx 0.75\%$.

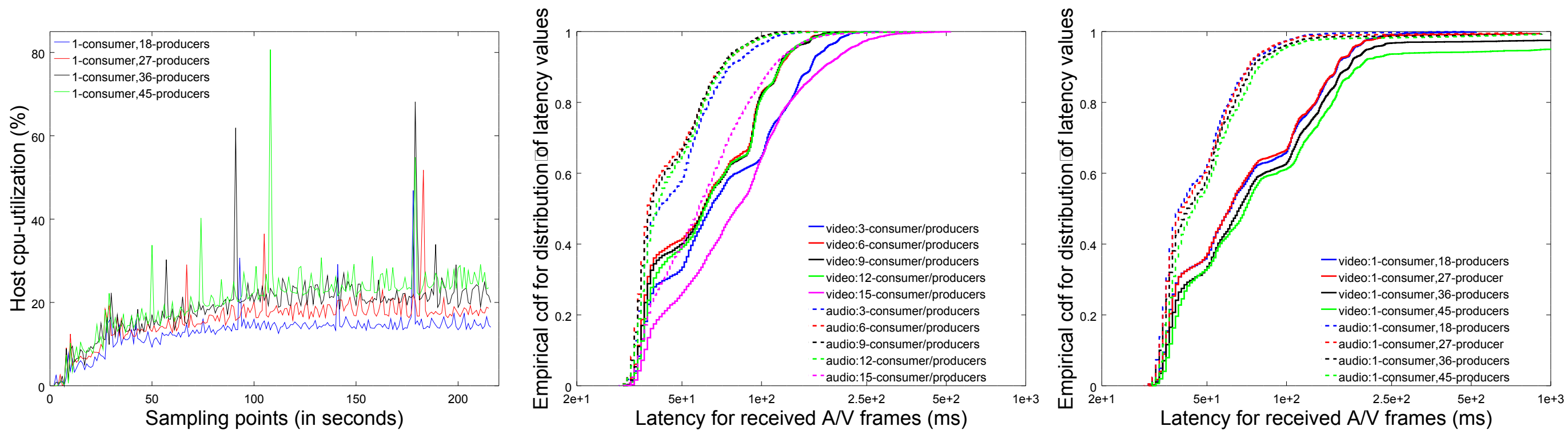

(a) CPU use at the consumer hosting server (b) Audio and video latency performance with all consumer-producer nodes (c) Audio and video latency performance with single consumer and multiple producer nodes our implementation of the underlying NDN forwarder to support the proposed architecture at the processing level, and (*ii*) the excessive number of java threads that are being used to fetch the content corresponding to the remote participants from the local forwarder. Further improvements are possible by taking better advantage of multi-threading or utilizing lock-free data structures, in addition to facilitating more efficient resource utilization schemes at the end hosts. However, such optimizations are currently outside the scope of our research.

Fig. 8. Performance concerning the scalability of the proposed architecture.

TABLE III
All Consumer-Producer Participants

| #Participants | 3 | 6 | 9 | 12 | 15 |
|---|---|---|---|---|---|
| $\text{Quality}_{\text{AUDIO}}$ (%) | 99.93 | 99.98 | 100.00 | 99.76 | 97.69 |
| $\text{Quality}_{\text{VIDEO}}$ (%) | 99.91 | 100.00 | 99.97 | 99.81 | 97.46 |

TABLE IV
Single-Consumer Multi-Producer Case

| #Participants | 19 | 28 | 37 | 46 |
|---|---|---|---|---|
| $\text{Quality}_{\text{AUDIO}}$ (%) | 99.51 | 99.01 | 98.53 | 97.93 |
| $\text{Quality}_{\text{VIDEO}}$ (%) | 99.16 | 98.91 | 96.78 | 93.63 |

We present the results on the perceived quality in Tables III and IV. The results correspond to the ratio of usable content at the consumer side, which takes into account lost and delayed frames (*i.e.*, frames arriving later than the decoding deadline). We observe acceptable performance for most of the cases. However, as the number of streams received by a host machine increases, we start to observe increased loss in performance, which suggest the host machine may become the bottleneck towards achieving a scalable architecture capable of supporting 100 or more participants.

### *E. Resiliency Results*

We illustrate, for a consumer-producer pair, the recovery performance by a consumer after it experiences a connectivity loss triggered by a link failure in Figure 9. Here, connectivity is considered from the perspective of the ICN layer, hence the down/up-times also include failure/recovery at the IP layer. In the figure, we show the arrival data rate at the consumer per 200ms intervals (or per 5-frame long periods), and the arrival status for the notifications at the consumer. We observe that, as soon as the consumer's link recovers (which occurs right before the notification arrives), the notification received from the sync-proxy along with the pre-fetching logic enables the consumer to successfully restore the video session, which for our current design takes 1-GOP period (or 1s) for recovery. It is possible to reduce this recovery latency by utilizing efficient coding schemes that utilize fast recovery frames.

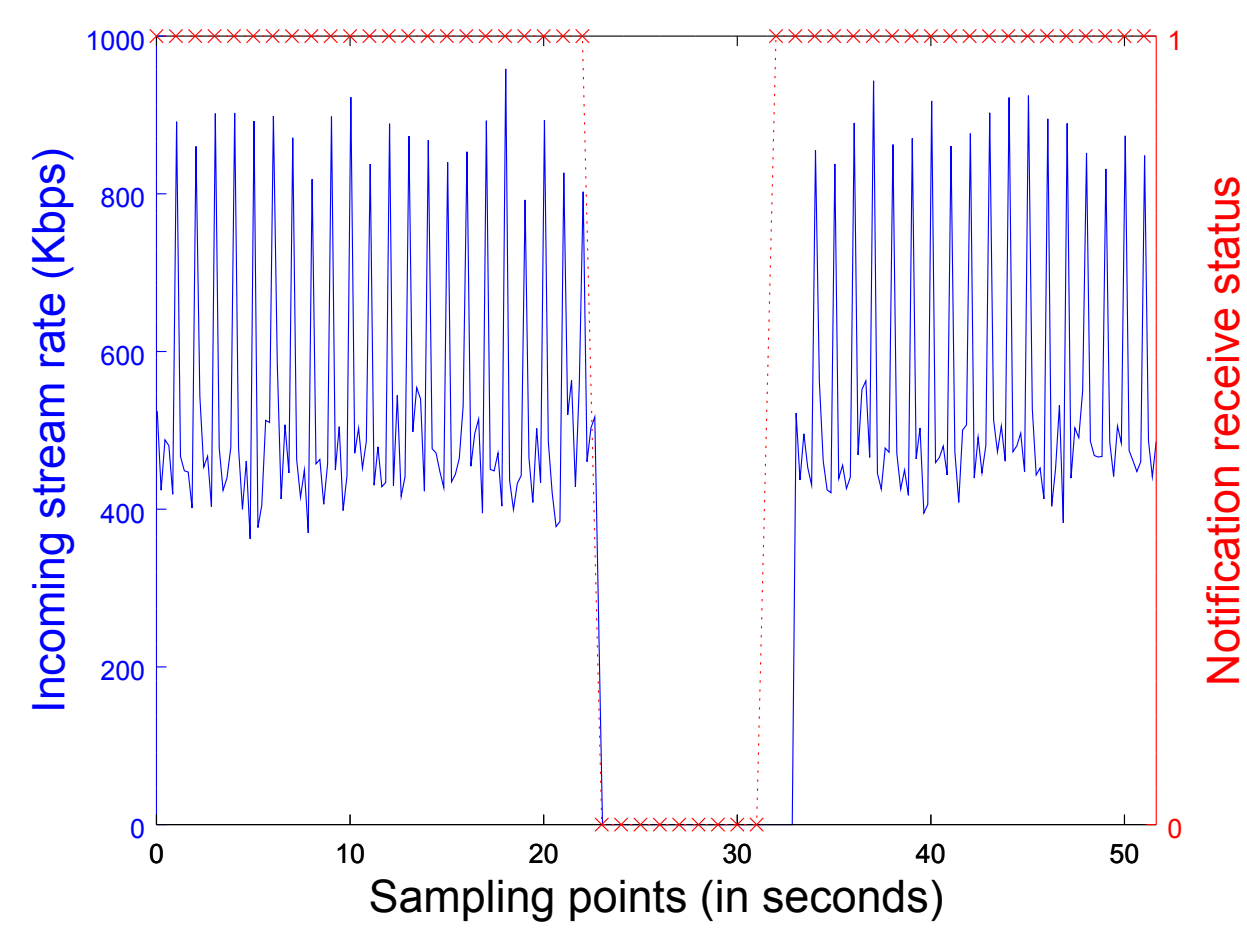

Fig. 9. Recovery at a consumer after connectivity loss.

## V. Related Work

VoCCN in [9], ACT in [22] and NDN-RTC in [10] and [23] have discussed the topic of real-time communication over ICN. The first two dealt with two-party and multi-party audio respectively, and established the feasibility of real-time communications, while pointing out the benefits in terms of scalability, robustness and security over similar IP based solutions. ACT placed emphasis on specific methods to achieve conference discovery and speaker discovery in order to generalize a distributed conferencing framework.

NDN-RTC addressed the additional complexities for generating, publishing, and consuming video content, and provided novel approaches to minimize the latency using a pull based communication framework. However, the application-support features offered by NDN-RTC continue to rely on heuristic approaches implemented at the end hosts, such as measuring response time for Interests during bootstrap phase or during connection disruptions for the wireless consumer. Conversely, our solution pushes many of these features towards the network as in-network services to reduce the end host complexity, which allows for better scalability performance.

The solution proposed in our paper is a result of evolution from the earlier works described in [13] and [12]. The former work in [13] focused on an edge-cloud based conferencing framework, and the latter work in [12] built upon the former's architecture to develop a CCN-based video conferencing solution. While our solution essentially provides similar benefits in terms of scalability, robustness and security as experienced by NDN-RTC and the other ICN-based real-time communication solutions, and experiences similar challenges of utilizing a pull-based architecture to support real-time communications; it uses a markedly different approach. SRMCA attempts to reduce the end-user complexity by offloading some of the conference framework functionality to the network. Critical operations like namespace synchronization are provided by the network as a service, resulting in a more deterministic response to join/leave events and network disruptions, and also opening up the case of a newer business model. A simplified end-user component allows higher scalability in terms of the number of participants as corroborated by our experimental analysis. This comes at the cost of losing the pure distributed nature of the application as certain functionalities are concentrated at network service points, but the network operator can mitigate that risk by introducing redundancy in the service framework.

## VI. Conclusion

In this paper, we proposed a scalable and resilient multi-party communication architecture capable of handling real-time communication challenges such as random join (or leave) events and fast session recovery by a participant after it experiences connectivity issues. To support quick and efficient name-syncing among participants, we employed an application agnostic notification framework that leveraged the service-friendly CCN transport. We provided an in-depth discussion of the proposed conferencing architecture and presented an analytical framework for its design and the evaluation of its components. We evaluated the performance of our solution using a realistic emulation-based implementation framework based on NDN and studied the key performance metrics including bandwidth, latency, and computing use at different components of the architecture. We showed the scalability of the proposed solution to handle close to 50 participants in our studies (and up to 100 participants based on approximations on the experimental results) demonstrating its capability to support a large-scale real-time multi-party communication framework, including but not limited to multi-party conferencing and multi-source streaming towards, for instance, efficient realization of applications such as multi-source mixed reality (MR).